\documentclass[11pt,english]{article}
\usepackage[T1]{fontenc}
\usepackage[latin9]{inputenc}
\usepackage{geometry}
\usepackage{hyperref}
\usepackage{cite}

\usepackage[usenames,dvipsnames]{color}
\usepackage{graphicx}
\usepackage{amsmath}
\usepackage{amsfonts}
\usepackage{amssymb}
\usepackage{dsfont}
\usepackage{dcolumn}
\usepackage{bm}
\usepackage{slashed}
\usepackage{slashed}
\usepackage{multirow}
\usepackage{array}
\usepackage{rotating}
\usepackage{xcolor}
\usepackage{epstopdf}
\usepackage{fancybox}
\usepackage{fancyvrb}
\usepackage[normalem]{ulem}
\usepackage{color}
\usepackage{mathdots}
\usepackage{mathrsfs}
\usepackage{multirow}
\usepackage{esint}
\usepackage{multirow,bigdelim}
\allowdisplaybreaks

\newcolumntype{x}[1]{
{\centering}p{#1}}%

\usepackage{amsfonts}
\usepackage{amssymb}
\usepackage{amsmath,amsfonts,amssymb}
\usepackage{epstopdf}
\usepackage[section]{placeins}
\usepackage{hyperref}
\usepackage{booktabs}
\usepackage{cleveref}
\crefname{equation}{Eq.}{Eqs.}
\crefname{figure}{Fig.}{Figs.}
\crefname{table}{Table}{Tables}
\crefname{section}{Section}{Sections}

\usepackage[usenames,dvipsnames]{color}
\usepackage{graphicx}
\usepackage{subfigure}
\usepackage{float}
\def\rmuu{\gamma^{\mu}}
\def\rmud{\gamma_{\mu}}
\def\PL{{1-\gamma_5\over 2}}
\def\PR{{1+\gamma_5\over 2}}
\def\sinW2{\sin^2\theta_W}
\def\AEM{\alpha_{EM}}
\def\mul{M_{\tilde{u} L}^2}
\def\mur{M_{\tilde{u} R}^2}
\def\mdl{M_{\tilde{d} L}^2}
\def\mdr{M_{\tilde{d} R}^2}
\def\mz2{M_{z}^2}
\def\c2b{\cos 2\beta}
\def\au{A_u}
\def\ad{A_d}
\def\cob{\cot \beta}
\def\v#1{v_#1}
\def\tb{\tan\beta}
\def\epem{$e^+e^-$}
\def\KK{$K^0$-$\overline{K^0}$}
\def\wi{\omega_i}
\def\xj{\chi_j}
\def\Wmu{W_\mu}
\def\Wnu{W_\nu}
\def\m#1{{\tilde m}_#1}
\def\mH{m_H}
\def\mw#1{{\tilde m}_{\omega #1}}
\def\mx#1{{\tilde m}_{\chi^{0}_#1}}
\def\mc#1{{\tilde m}_{\chi^{+}_#1}}
\def\mwi{{\tilde m}_{\omega i}}
\def\mxi{{\tilde m}_{\chi^{0}_i}}
\def\mci{{\tilde m}_{\chi^{+}_i}}

\def\ch{{\tilde\chi^{+}_1}}
\def\c2{{\tilde\chi^{+}_2}}

\def\tt{{\tilde\theta}}

\def\tp{{\tilde\phi}}

\def\mz{M_z}
\def\sw{\sin\theta_W}
\def\cw{\cos\theta_W}
\def\cb{\cos\beta}
\def\sb{\sin\beta}
\def\rwi{r_{\omega i}}
\def\rxj{r_{\chi j}}
\def\rfp{r_f'}
\def\Kik{K_{ik}}
\def\Fq2{F_{2}(q^2)}
\def\f{\({\cal F}\)}
\def\d1{{\f(\tilde c;\tilde s;\tilde W)+ \f(\tilde c;\tilde \mu;\tilde W)}}
\def\tw{\tan\theta_W}
\def\sec2w{sec^2\theta_W}
\begin{document}
\baselineskip 18pt
\def\today{\ifcase\month\or
 January\or February\or March\or April\or May\or June\or
 July\or August\or September\or October\or November\or December\fi
 \space\number\day, \number\year}
\def\thebibliography#1{\section*{References\markboth
 {References}{References}}\list
 {[\arabic{enumi}]}{\settowidth\labelwidth{[#1]}
 \leftmargin\labelwidth
 \advance\leftmargin\labelsep
 \usecounter{enumi}}
 \def\newblock{\hskip .11em plus .33em minus .07em}
 \sloppy
 \sfcode`\.=1000\relax}
\let\endthebibliography=\endlist
\def\lsim{\ ^<\llap{$_\sim$}\ }
\def\gsim{\ ^>\llap{$_\sim$}\ }
\def\r2{\sqrt 2}
\def\beq{\begin{equation}}
\def\eeq{\end{equation}}
\def\beqn{\begin{eqnarray}}
\def\eeqn{\end{eqnarray}}
\def\rmuu{\gamma^{\mu}}
\def\rmud{\gamma_{\mu}}
\def\PL{{1-\gamma_5\over 2}}
\def\PR{{1+\gamma_5\over 2}}
\def\sinW2{\sin^2\theta_W}
\def\AEM{\alpha_{EM}}
\def\mul{M_{\tilde{u} L}^2}
\def\mur{M_{\tilde{u} R}^2}
\def\mdl{M_{\tilde{d} L}^2}
\def\mdr{M_{\tilde{d} R}^2}
\def\mz2{M_{z}^2}
\def\c2b{\cos 2\beta}
\def\au{A_u}
\def\ad{A_d}
\def\cob{\cot \beta}
\def\v#1{v_#1}
\def\tb{\tan\beta}
\def\epem{$e^+e^-$}
\def\KK{$K^0$-$\bar{K^0}$}
\def\wi{\omega_i}
\def\xj{\chi_j}
\def\Wmu{W_\mu}
\def\Wnu{W_\nu}
\def\m#1{{\tilde m}_#1}
\def\mH{m_H}
\def\mw#1{{\tilde m}_{\omega #1}}
\def\mx#1{{\tilde m}_{\chi^{0}_#1}}
\def\mc#1{{\tilde m}_{\chi^{+}_#1}}
\def\mwi{{\tilde m}_{\omega i}}
\def\mxi{{\tilde m}_{\chi^{0}_i}}
\def\mci{{\tilde m}_{\chi^{+}_i}}
\def\mz{M_z}
\def\sw{\sin\theta_W}
\def\cw{\cos\theta_W}
\def\cb{\cos\beta}
\def\sb{\sin\beta}
\def\rwi{r_{\omega i}}
\def\rxj{r_{\chi j}}
\def\rfp{r_f'}
\def\Kik{K_{ik}}
\def\Fq2{F_{2}(q^2)}
\def\f{\({\cal F}\)}
\def\d1{{\f(\tilde c;\tilde s;\tilde W)+ \f(\tilde c;\tilde \mu;\tilde W)}}
\def\tw{\tan\theta_W}
\def\sec2w{sec^2\theta_W}
\def\ch{{\tilde\chi^{+}_1}}
\def\c2{{\tilde\chi^{+}_2}}

\def\tt{{\tilde\theta}}

\def\tp{{\tilde\phi}}

\def\mz{M_z}
\def\sw{\sin\theta_W}
\def\cw{\cos\theta_W}
\def\cb{\cos\beta}
\def\sb{\sin\beta}
\def\rwi{r_{\omega i}}
\def\rxj{r_{\chi j}}
\def\rfp{r_f'}
\def\Kik{K_{ik}}
\def\Fq2{F_{2}(q^2)}
\def\f{\({\cal F}\)}
\def\d1{{\f(\tilde c;\tilde s;\tilde W)+ \f(\tilde c;\tilde \mu;\tilde W)}}

\def\b{$\cal{B}(\tau\to\mu \gamma)$~}

\def\st{Stueckelberg~extension~}

\def\be{\begin{equation}}
\def\ee{\end{equation}}

\def\GeV{{\rm GeV}}
\newcommand{\bea}{\begin{eqnarray}}
\newcommand{\eea}{\end{eqnarray}}

\newcommand{\non}{\nonumber \\}

\def\tw{\tan\theta_W}
\def\sec2w{sec^2\theta_W}
\newcommand{\pxn}[1]{{\color{red}{#1}}}
\newcommand{\zl}[1]{{\color{blue}{#1}}}
\newcommand{\nc}[1]{{\color{green}{#1}}}

\def\a{\alpha}
\def\as{\alpha_s}
\def\b{\beta}
\def\c{\gamma}
\def\d{\delta}
\def\e{\epsilon}
\def\f{\varepsilon}

\def\mA{\mathcal{A}}
\def\mB{\mathcal{B}}
\def\mC{\mathcal{C}}
\def\mD{\mathcal{D}}
\def\mE{\mathcal{E}}
\def\mF{\mathcal{F}}
\def\mG{\mathcal{G}}
\def\mH{\mathcal{H}}
\def\mI{\mathcal{I}}
\def\mJ{\mathcal{J}}
\def\mK{\mathcal{K}}
\def\mL{\mathcal{L}}
\def\mM{\mathcal{M}}
\def\mN{\mathcal{N}}
\def\mO{\mathcal{O}}
\def\mP{\mathcal{P}}
\def\mQ{\mathcal{Q}}
\def\mR{\mathcal{R}}
\def\mS{\mathcal{S}}
\def\mT{\mathcal{T}}
\def\mU{\mathcal{U}}
\def\mV{\mathcal{V}}
\def\mW{\mathcal{W}}
\def\mX{\mathcal{X}}
\def\mY{\mathcal{Y}}
\def\mZ{\mathcal{Z}}

\begin{titlepage}

\begin{center}
{\large {\bf 
3.5 keV Galactic  Emission Line as  a Signal from the Hidden Sector
}}\\

\vskip 0.5 true cm
 Ning Chen$^{a}$\footnote{Email: ustc0204.chenning@gmail.com},
  Zuowei Liu$^{a,b}$\footnote{Email: zuoweiliu@tsinghua.edu.cn}
  and Pran Nath$^{c}$\footnote{Emal: nath@neu.edu}
\vskip 0.5 true cm
\end{center}
\noindent
{$^{a}$ Institute of Modern Physics and Center for High Energy Physics, Tsinghua University, Beijing, 100084,
China}\\
{$^{b}$ Department of Physics, McGill University, 3600 Rue University, Montreal, Quebec, Canada H3A 2T8 }\\
{$^{c}$ Department of Physics, Northeastern University, Boston, Massachusetts  02115-5000, USA} \\

\vskip 0.5 true cm

\centerline{\bf Abstract}
An emission line with energy of $E\sim 3.5$ keV has been observed in galaxy clusters by two 
experiments.  The emission line is consistent with the decay of a dark matter particle with a mass of
$\sim 7$ keV.  
In this work we discuss the possibility that the dark particle responsible for the 
emission is a real  scalar ($\rho$) which arises naturally in a $U(1)_X$ \st  of MSSM.
In the MSSM \st  $\rho$ couples only to other scalars carrying a $U(1)_X$ quantum number. 
Under the assumption that there exists a vectorlike leptonic generation 
carrying both $SU(2)_L\times U(1)_Y$ and $U(1)_X$ quantum numbers,
we compute the decay of the $\rho$ into two 
photons via a triangle loop involving scalars.
The relic density of the $\rho$ arises via   the decay  $H^0\to h^0+ \rho$ 
at the loop level involving scalars,
 and via the  annihilation  processes of the vectorlike scalars  into $\rho + h^0$. 
It is shown that the galactic data can be explained within a multicomponent dark matter model where the 
7 keV dark matter is a subdominant component constituting only 
$(1-10)$\%
of the matter relic density 
with the rest being supersymmetric dark matter such as the neutralino. Thus the direct detection
experiments remain viable searches for WIMPs. The fact that the dark scalar $\rho$ with no interactions with
the standard model particles  arises from a \st  of 
a hidden $U(1)_X$ implies that the  3.5 KeV galactic line emission is a signal from the hidden sector.

 \noindent
{\scriptsize
PACS numbers:}

\medskip

\end{titlepage}
\section{Introduction \label{sec1}}

Two experiments ~\cite{Bulbul:2014sua,Boyarsky:2014jta} have
seen a 3.5 keV gamma  line in the sky.  A possible explanation is that it is coming from 
decay of a dark matter particle.  
The experiment  gives
   \be
  \Gamma^{-1}(\rho \to \gamma\gamma) =(4\times 10^{27}- 4\times 10^{28})s \ . 
  \label{1}
  \ee 
   Several works already exist trying to explain the 3.5 keV line 
   such as from decaying dark particle
   \cite{Babu:2014pxa} and emission from a dark atom~\cite{Frandsen:2014lfa,Cline:2014eaa}.   
   Here we consider the possibility that the gamma  emission arises from the decay of a real field $\rho$ that arises
   naturally in a $U(1)_X$ extension of the standard model gauge group 
\cite{Kors:2004dx,Kors:2004ri,Kors:2004iz,Kors:2005uz} (see also \cite{Cheung:2007ut,Feldman:2007nf,Feldman:2007wj,Feldman:2006wb,
Liu:2011di,Feldman:2011ms,Feng:2012jn,Perez:2014gta,Santos:2014xka}).
In the $U(1)_X$ \st of MSSM we assume that all the MSSM particles are neutral under $U(1)_X$ but
that there exists extra matter in the form of a vectorlike multiplet which is charged under $U(1)_X$. 
As mentioned in such an extension there  naturally exists a real scalar particle $\rho$ which couples only to complex scalar particles
which are charged under $U(1)_X$. In this work we will assume 
that extra matter consists of vectorlike multiplets which  transform under $SU(2)_L\times U(1)_Y$ as doublets and singlets. 
We assume that $\rho$  has a mass of $\sim 7$ keV and it decays  to two photons via exchange of these
charged scalar fields 
to produce the  galactic 3.5 keV gamma  line.  Regarding relic density,
we assume that  the primordial relic density is inflated away and the current relic density arises from the decay of  Higgs
bosons, and also scalar annihilations. There are many processes that can contribute to the relic density.  These are: $h^0\to \rho + \rho$, $H^0\to \rho  + \rho$,
$H^0\to \rho + h^0$, 
{$\tilde E_1 \bar{ \tilde E}_1\to \rho +\rho, \rho + h^0$} where $E_1$ is a charged lepton in the vectorlike
multiplet  etc.
  It turns out that the final states
 with $\rho \rho$ are suppressed by $m_\rho^4$  and are thus negligible.  Further, the
 contribution of the Higgs boson decays dominate the contribution from the annihilation 
 to the relic density. Within the above model it is found possible to fit the galactic data with the 7 keV scalar dark
 matter being a subdominant component making up as little as 
1\%-10\% of the total dark matter density with the
 rest being supersymmetric dark matter such as the neutralino or the gravitino.\\

The outline of the rest of the paper is as follows: In \cref{sec2} we give a brief description of the 
\st of MSSM with inclusion of vectorlike multiplets which are charged both under $SU(2)_L\times U(1)_Y$
and under $U(1)_X$.  
 In \cref{sec3} we give an analysis of 
 lifetime of the $\rho$ decaying into two photons via triangle loops involving scalars charged under 
 $U(1)_X$ and also charged under $ U(1)_{\rm em}$. In \cref{sec4} we give an analysis of the 
 relic density of the $\rho$ which is produced via the decay of the Higgs bosons and via annihilation
 processes. Here we show that the relic density of the $\rho$ arising from Higgs decay is the 
 dominant component and the annihilation processes are subdominant. In \cref{sec5} we give
 a numerical analysis where we fit the gamma  data from the galactic clusters
 with $\rho$ as a subdominant component of dark matter.  Conclusions are 
 given in {\cref{sec6}} while further details of the analysis are given in the appendices.

\section{The Model\label{sec2}}
As mentioned in Section\ \ref{sec1}
we consider an extra vectorlike leptonic generation $V$ consisting of 
{$L, E^c, N^c, L^{\prime c}, E', N'$} 
with $SU(3)_C\times SU(2)_L\times U(1)_Y\times U(1)_X$ quantum numbers as follows

\begin{align}\label{2}
&L= \left(\begin{matrix}  N_{L}\cr
 ~{E}_{L}  \end{matrix} \right)  (1,2,- \frac{1}{2},1),
&& E^c_{L}(1,1,1, -1),
& & N^c_{L}(1,1,0,-1)\ ,\\
&L^{{\prime c}} =\left(\begin{matrix} E_{ L}^{{\prime c}} \cr
 N_L^{{\prime c}}\end{matrix}\right)(1,2,\frac{1}{2}, -1),
&& E_L' (1,1,-1,1), 
 &&  N_L'(1,1,0,1)\ ,
\label{3}
\end{align}
where the first two numbers refer to the representation.
The scalar fields can be written as above with a tilde on them except that 
$\tilde E^c_L= \tilde E_R^*,  \tilde E^{{\prime c}}= \tilde E_R^{{\prime *}}$.
We also here  note that the Higgs doublets in the MSSM have the quantum numbers
\begin{gather}
H_d=(\mathbf{1},\mathbf{2},-\tfrac{1}{2},0)\,,\qquad H_u=(\mathbf{1},\mathbf{2},+\tfrac{1}{2},0)\, \ .
\end{gather}
{Here we also note that all of the MSSM particles carry no $U(1)_X$ charge.}
The superpotential for the vectorlike leptonic supermultiplets is given by
{
\begin{multline}
W = yLH_{d}E_L^{c}+y'L^{{\prime c}}H_{u}E_L'+M_{L}LL^{{\prime c}}+M_{E}E_L'E_L^{c}+ M_N N_L'N_L^c
\, ,
\label{yuk}
\end{multline}
}
where $M_{L}$, $M_{E}$ and $M_N$ are the vectorlike masses.  
After spontaneous breaking the two Higgs doublets of $SU(2)_L$  develop VEVs so that:
\beqn
H_d = \left(\begin{matrix} H_d^0\cr
                           H_d^-\end{matrix}\right)
= \left(\begin{matrix}
             \frac{1}{\sqrt 2}(v_d+\phi_1)\cr
             H_d^-\end{matrix}\right)\,,\qquad
H_u= \left(\begin{matrix}H_u^+\cr
             H_u^0\end{matrix}\right)
=\left(\begin{matrix}H_u^+ \cr
            \frac{1}{\sqrt 2}(v_u+\phi_2) \end{matrix}\right)\,,
\eeqn
where $v_d$ and $v_u$ are the VEVs of $H^0_d$ and $H^0_u$.
Now $\rho$ couples only to the scalars so we focus on the scalar fields which are charged under
$U(1)_X$.  In this case we have a 
$4\times 4$ mass {squared} matrix and in  the basis {$(\tilde E_L, \tilde E_R, \tilde E_L', \tilde E_R')$}
it is given by
\footnote{{ In the analysis below the $2\times 2$ off -diagonal matrices in \cref{slep1} 
will be neglected. They are displayed in \cref{slep1} for completeness. }}

\begin{equation}
\frac{1}{\sqrt{2}}
\left(\begin{array}{cc|cc}
\multicolumn{2}{c|}{\multirow{2}*{$\sqrt{2}(M_{\tilde{E}}^2)_{2\times 2}$}} & y' v_u M_L + y v_d M_E & 0\\
&& 0 & y' v_u M_E + y v_d M_L \\
\hline
y' v_u M_L + y v_d M_E & 0 & \multicolumn{2}{c}{\multirow{2}*{$\sqrt{2} (M_{\tilde{E}'}^2)_{2\times 2}$}}\\
0 & y' v_u M_E + y v_d M_L &&\\
\end{array}\right)_{4\times 4}\,.
\label{slep1}
\end{equation}
Here $(M^2_{\tilde E})_{2\times 2}$ is given by
\beqn
(M^2_{\tilde E})_{2\times 2}=\left(\begin{array}{cc}
M_{1}^{2}+\tfrac{1}{2}y^2 v^2_d +M_{L}^{2} +\frac{(g_1^2-g_2^2)}{8} (v_d^2 - v_u^2)
& \frac{1}{\sqrt 2} y (A_E v_d  - \mu v_u)\\
\frac{1}{\sqrt 2} y (A_{E} v_d  - \mu v_u)
&   M_1^2 +\tfrac{1}{2}y^2 v^2_d+ M_E^2 - \frac{g_1^2}{4} (v_d^2 - v_u^2)
\end{array}\right)\,,
\label{slep2}
\eeqn
where $M_1$ is the soft mass 
while $M_L$ and $M_E$ are vectorlike masses. We label the eigenvalues
as 
$m^2_{\tilde E_1}$ and $m^2_{\tilde E_2}$  
and the corresponding eigenstates by $\tilde E_1$ and $\tilde E_2$ which are related
to $\tilde E_L$ and $\tilde E_R$ by 
\begin{gather}
  \left(\begin{matrix} \tilde E_L \\ 
\tilde E_R\end{matrix}\right)
= \left(\begin{matrix} \cos\xi & \sin\xi \\
-\sin\xi & \cos\xi \end{matrix} \right)
\left(\begin{matrix} \tilde E_1 \\ 
\tilde E_2\end{matrix}\right)\ . 
 \end{gather}
Similarly  $(M^2_{\tilde E'})_{2\times 2}$ is given by
\beqn
(M^2_{\tilde E'})_{2\times 2}
=\left(\begin{array}{cc}
M_{2}^{2}+\tfrac{1}{2}y'^2 v^2_u+M_{L}^{2} -\frac{(g_1^2-g_2^2)}{8} (v_d^2 - v_u^2)
& \frac{1}{\sqrt 2} y' (A_{E'} v_u  - \mu v_d)\\
\frac{1}{\sqrt 2} y' (A_{E'} v_u  - \mu v_d)
&   M_2^2 +\tfrac{1}{2}y'^2 v^2_u+ M_E^2 +\frac{g_1^2}{4} (v_d^2 - v_u^2)
\end{array}\right)\, 
\label{slep3}
\eeqn
where $M_2$ is the soft mass. 
 We label the eigenvalues of this mass squared matrix by 
$m^2_{\tilde E'_1}, m^2_{\tilde E'_2}$ 
 with the corresponding eigenstates as $\tilde E_1'$ and $\tilde E_2'$. They are related to 
$\tilde E_L'$  and $\tilde E_R'$ by 
  \begin{gather}
  \left(\begin{matrix} \tilde E'_L \\ 
\tilde E'_R\end{matrix}\right)
= \left(\begin{matrix} \cos\xi' & \sin\xi' \\
-\sin\xi' & \cos\xi' \end{matrix} \right)
\left(\begin{matrix} \tilde E_1' \\ 
\tilde E_2'\end{matrix}\right)\ . 
 \end{gather}
Since the diagonalization of the $4\times 4$ scalar mass squared matrix is intractable, we consider the 
case where 
the product of the fermion masses and
the vectorlike masses are much smaller than the soft 
mass squares. 
In this case the mass squared matrix of 
\cref{slep1}  takes on a block diagonal form where the $2\times 2$ matrix in the upper left hand corner 
 is the mass squared matrix for  the normal leptons in the vectorlike multiplet and 
 the $2\times 2$ matrix in the lower right hand corner  is the mass squared matrix for the mirror leptons in the vector 
 like multiplet. \\

{
 In the analysis we have a $U(1)_X$ extension of MSSM and we assume that  $U(1)_X$ gauge boson
 acquires mass through a Stueckelberg mechanism.  This mechanism works only for $U(1)$ extensions. 
 To achieve the extension we need to include a vector superfield $C$ and scalar supefields
 $S$ and $\bar S$ and assume a Lagrangian of the type 

%
\beqn
{\cal L}_{\rm St} = \int d\theta^2 d\bar{\theta}^2\, \left[ M_C C + S +\bar S \right]^2\ , 
\label{mass}
\eeqn 
%
The Lagrangian is invariant under the $U(1)_X$ gauge transformations
\beqn \label{stgauge}
\delta_X C = \Lambda_X + \bar\Lambda_X\ , \quad \delta_X S = - M_C \Lambda_X\ . 
\eeqn
%
The vector superfield  $C$ in Wess-Zumino gauge is given by   
\beqn
C= -\theta\sigma^{\mu}\bar \theta C_{\mu} 
+i\theta\theta \bar\theta \bar \lambda_C  
-i\bar\theta\bar\theta \theta  \lambda_C
+\frac{1}{2} \theta \theta\bar\theta\bar\theta D_C \ , 
\eeqn 
while  $S$ is given by 
\beqn \label{supS}
S
&=&\frac{1}{2}(\rho +ia ) + \theta\chi 
 + i \theta\sigma^{\mu}\bar\theta \frac{1}{2}(\partial_{\mu} \rho 
+i \partial_{\mu} a) 
\\
&&
+~ \theta\theta F +\frac{i}{2} \theta \theta \bar\theta \bar\sigma^{\mu} \partial_{\mu} \chi 
+\frac{1}{8}\theta\theta\bar\theta\bar\theta (\Box \rho+i\Box a) \ . 
\nonumber
\eeqn
The  complex scalar component of $S$ 
 contains the axionic pseudo-scalar $a$ and a real scalar field $\rho$. 
${\cal L}_{\rm St}$ in component 
notation takes the form
\beqn \label{stueck} 
{\cal L}_{\rm St} &=& - \frac{1}{2}(MC_{\mu}  +\partial_{\mu} a)^2
- \frac{1}{2} (\partial_\mu \rho)^2 
- i \chi \sigma^{\mu} \partial_{\mu}\bar {\chi} + 2|F|^2 
\nonumber\\
&& 
\hspace{0cm}
+M\rho  D_C
+M\bar {\chi} \bar \lambda_C
 + M\chi  \lambda_C \ . 
\label{Lst}
\eeqn
For the gauge field we add the kinetic terms
\beqn 
{\cal L}_{\rm gkin} = 
-\frac{1}{4}  C_{\mu\nu} C^{\mu\nu} 
- i \lambda_C \sigma^{\mu}\partial_{\mu} \bar\lambda_C
+\frac{1}{2} D_C^2  
\label{Lkin}
\eeqn 
where $C_{\mu\nu}=\partial_\mu C_\nu - \partial_\nu C_\mu$.
For the matter fields (i.e., hidden sector matter) 
chiral superfields  with components $(f_i,z_i,F_i)$ are introduced, which are defined similar to 
$S$.
Their Lagrangian is standard and we do not display it here.  It is the real
scalar field $\rho$  which is the focus of our study here.
From Eq.(\ref{Lst}) and Eq.(\ref{Lkin}) it  is seen that the mass of $C_\mu$ which is identified
in the unitary gauge as $Z'$ is the same as the mass of $\rho$ which can be gotten by 
elimination of the auxiliary field $D_C$. \\
}

\begin{figure}[htbp]
\begin{center}
\includegraphics[width=0.5\columnwidth]{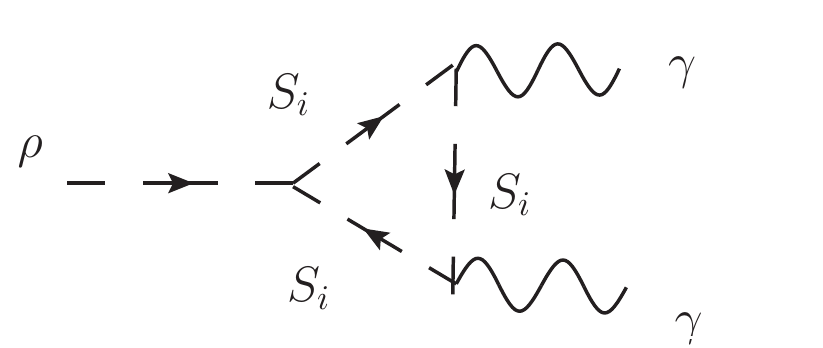}
\caption{Triangle loop  diagram for the decay process $\rho \to \gamma\gamma$ 
 via exchange of vectorlike scalars  (arising from  \cref{2} and \cref{3}) 
  in the loop which are charged under $U(1)_X$ and
 under $U(1)_{\rm em}$.
}
\label{fig:feynlife}
\end{center}
\end{figure}

\section{$\rho$  lifetime\label{sec3}}
We use the interactions of \cref{sec7} to compute the $\rho$ lifetime.
The decay width of $\rho$ through scalar loops is given by (see, e.g., \cite{Feng:2013mea})
\be
\Gamma(\rho \to \gamma\gamma) = \frac{\alpha^2 m_{\rho}^3}{1024 \pi^3} 
\left[ \sum_i\frac{g_{\rho S_iS_i}}{m_{S_i}^2} N_{C,S} q_{S_i}^2 A_0(\tau_{s_i})\right]^2 \ , 
\label{width}
\ee
 Here the explicit form of $g_{\rho S_iS_i}$, where $S_i$ are in the mass diagonal basis, 
is given in 
{\cref{sec7}.} 
In \cref{width} $N_{C,S}$ are the (color, spin) multiplicities and  
$q_{S_i}$ is the electric charge for the field $S_i$ under $U(1)_{\rm em}$.
We note that $\rho$ does not couple with the particles in the standard model, i.e., with quarks and leptons, 
or with the Higgs  or with $W^{\pm}$ so these
particles cannot appear in the loop. Only the scalars charged under $U(1)_X$ and under $U(1)_{\rm em}$ 
can  appear in the loop.
In \cref{width}  
$\alpha= 1/137, m_{\rho}= 7 ~\text{keV}= 7\times 10^{-6}~\GeV$, 
where $\tau_s$ is defined by 
$\tau_s= \frac{4 m_S^2}{m_{\rho}^2}$. 
 $A_0(\tau)$ that appears in Eq.(\ref{width}) 
 is a loop function which is given by 
\be
A_0(\tau)=  -\tau [ 1- \tau f(\tau)]\  , 
\ee
where $f(\tau)$ is defined by
\begin{equation}
f(\tau)=
\left\{
\begin{aligned} & \Big( \arcsin \frac{1}{\sqrt \tau} \Big)^{2}\,,\qquad\tau\geq1\, \  , \\
 & -\frac{1}{4}\Big[\ln\frac{\eta_{+}}{\eta_{-}}-i\pi\Big]^{2}\,,\qquad\tau<1\,,
\end{aligned}
\right.
\end{equation}
where $\eta_{\pm}\equiv (1 \pm \sqrt{1-\tau})$ and
$\tau = 4 m^2 \big/ m_\rho^2$ for a particle running in the loop with mass $m$.
For the case when $\tau \gg 1$ one has
\beqn
f(\tau)\to \frac{1}{\tau}(1+ \frac{1}{3\tau} +
 \cdots )\,,
\eeqn
and in this limit  $A_0\to 1/3$. For our case $\tau\gg1$ so we can replace $A_0$ by $1/3$. 
 Next using the result of 
{\cref{sec7}} we have
\be
\sum_i\frac{g_{\rho S_iS_i}}{m_{S_i}^2}  N_{C,S} q_{S_i}^2 A_0(\tau_{s_i})
=g_Xq_E^2Q_E  \cos 2\xi  \frac{m_{\rho}}{3m_{S}^2}   
+g_Xq_{E'}^2 Q_{E'}  \cos 2\xi'  \frac{m_{\rho}}{3m_{S'}^2} \  .
\ee
Here  we have  set $A_0=1/3$, and 
$m_S^2$ and $m_{S'}^2$
are effective scalar mass squares defined by 
\begin{gather}
m_S^2= \frac{m_{\tilde E_1}^2 m_{\tilde E_2}^2}{m_{\tilde E_2}^2-m_{\tilde E_1}^2},
~~m_{S'}^2= \frac{m_{\tilde E_1'}^2 m_{\tilde E_2'}^2}{m_{\tilde E_2'}^2-m_{\tilde E_1'}^2}\ , 
\end{gather}
where $m_{\tilde E_1}$ and $m_{\tilde E_2}$ are mass eigenvalues corresponding to the eigenstates $\tilde E_1$ and
$\tilde E_2$ etc. The width formula now simples to 
\be
\Gamma(\rho \to \gamma\gamma) = \frac{\alpha^2 m_\rho^5 g_X^2}{9\times 1024 \pi^3} 
\left(\frac{  Q_E\cos 2\xi}{m_S^2} + \frac{Q_{E'}\cos 2\xi'}{m_{S'}^2}\right)^2\  , 
\ee 
where we have used $q_E=  q_{E'}= 1$.
A numerical analysis of $\Gamma(\rho \to \gamma\gamma)$ along with the relic density constraint
will be discussed in {\cref{sec5}.} 

\section{Relic density analysis\label{sec4}}

\begin{figure}[htbp]
\begin{center}
\includegraphics[width=0.5\columnwidth]{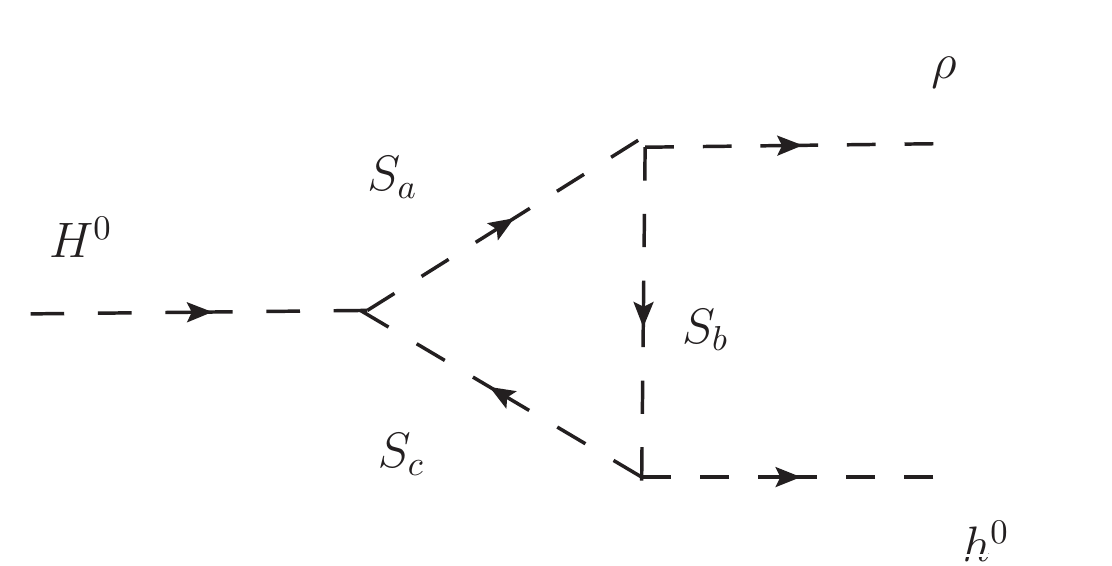}
\caption{
Triangle loop  diagram for the decay process $H^0 \to \rho + h^0$ 
 via exchange of vectorlike scalars (arising from  \cref{2} and \cref{3}) 
 in the loop which are charged under $U(1)_X$.  
}
\label{fig:feynRD1}
\end{center}
\end{figure}

\subsection{The decay $H^0 \to h^0 + \rho$  }
\label{sec:rddecay}

The $H^0$ decays into $h^0+ \rho$ via triangle loops and there are 
$4^3=64$ 
triangle loops to consider. 
Here we neglect the mixing due to the off diagonal terms in Eq.\ (\ref{slep1}), 
and only consider 16 different triangle diagrams.
{The reduction {from} 64 to 16 is due to the neglect of the off diagonal terms in Eq.\ (\ref{slep1}).}
Thus in a compact notation we can write the interaction of $\rho$ 
with the scalars charged under $U(1)_X$ as 

\be
{\cal L}_{\text{st}}=  m_\rho g_X Q_E \rho 
g_{\rho ij}
\tilde E_i^* \tilde E_j 
+  m_\rho g_X Q_{E'} \rho 
g_{\rho ij}'
\tilde E_i^{{\prime *}} \tilde E_j' \  , 
 ~~~~~i, j=1,2
\ee
where $g_{\rho ij}$ and $g_{\rho ij}'$
are the 
 ``reduced'' couplings and are given by (along with other couplings to the Higgses)
{
\bea
g_{\rho 1 1} = - g_{\rho 2 2} = g_{h 1 2} = g_{h 2 1} = g_{H 1 2} = g_{H 2 1} = \cos 2\xi \ , \\
g_{\rho 1 2} = g_{\rho 2 1} = -g_{h 1 1} = g_{h 2 2} = -g_{H 1 1} = g_{H 2 2} = \sin2\xi \  , \\
g'_{\rho 1 1} = - g'_{\rho 2 2} = g'_{h 1 2} = g'_{h 2 1} = g'_{H 1 2} = g'_{H 2 1} = \cos 2\xi' \ , \\
g'_{\rho 1 2} = g'_{\rho 2 1} = -g'_{h 1 1} = g'_{h 2 2} = -g'_{H 1 1} = g'_{H 2 2} = \sin2\xi' \  .
\eea
}
Let us consider the $g_{ij}$ couplings. Since each scalar propagator can be either $\tilde E_1$ or $\tilde E_2$, 
there are eight different triangle diagrams that contribute.  
For a specific diagram  labeled by the three
scalars in the loop as $(a,b,c)$, as shown in Fig.\ (\ref{fig:feynRD1}), the matrix element is given by 
\be
{\cal M}_{abc} = G_0 ~  g_{\rho ab} g_{h bc} g_{Hca} \, 
I_{abc}
\ee
where 
\be
G_0 \equiv  m_\rho g_X Q_E \left( {g_2 m_E \over 2 {M_W} \cos\beta} \right)^2 
(-A_E \sin\alpha + \mu \cos\alpha)
(A_E \cos\alpha + \mu \sin\alpha)\  , 
\ee
and 
\be
I_{abc} \equiv \int {d^4 k  \over (2\pi)^4} \left[ 
{1 \over (k+p_2)^2-m_a^2}  
{1 \over k^2-m_b^2} 
{1 \over (k-p_3)^2-m_c^2}
\right]\ . 
\ee
Using Feynman parameterization, we obtain the loop integral as 
\be
I_{abc}  = {-i \over (4\pi)^2} {1\over B_2}
\int_0^1 dx 
\left[
f\left({B_1 \over 2 B_2} + 1-x, A  \right)
-f\left({B_1 \over 2 B_2},  A  \right)
\right] \  , 
\ee
where 
\bea
B_0 &\equiv& x^2 m_\rho^2 + x(m_a^2 -m_b^2-m_\rho^2) + m_b^2 \  , \\
B_1 &\equiv& m_c^2 -m_b^2 - m^2_{h^0}  -x(m^2_{H^0} - m_\rho^2 - m^2_{h^0}) \  , \\
B_2 &\equiv& m^2_{h^0} \  ,  \\
{A} &\equiv& {4 B_0 B_2 -B_1^2 \over 4 B_2^2} \  , 
\eea
and the function $f(x, A)$ is defined as 
\be
f(x,A) \equiv {1 \over \sqrt{|A|}}  \left\{
 \begin{array}{cc} 
 \arctan(x/\sqrt{A}); & A >0 \\ 
\ln\sqrt{x-\sqrt{-A} \over x+\sqrt{-A}}; & A <0  \  . \end{array}
\right. 
\ee
The total matrix element is then given by 
\be
{\cal M} = G_0 \sum_{a,b,c} ( g_{\rho a b} ~ g_{hbc}~ g_{H ca} ~ I_{abc})\ . 
\ee
Summing over all possibilities, we have 
\bea
{\cal M} /G_0  =c^3(I_{112}-I_{221}) 
+cs^2(I_{111}-I_{121}+I_{122}
-I_{211}+I_{212}-I_{222})\  , 
\eea
where $c\equiv \cos(2\xi)$ and 
$s\equiv  \sin(2\xi)$. 
The decay width of the process $H^0 \to h^0 + \rho$ can now  be 
computed  

\be
\Gamma(H^0 \to h^0 + \rho) = 
{1 \over 16 \pi m_{H^0}}
\left[1- \left({m_{h^0} \over m_{H^0}}\right)^2\right]
\overline{|{\cal M  + \cal M'}|^2} \ . 
\ee
where ${\cal M}'$ is the amplitude with $g_{ij}$ replaced by $g'_{ij}$, 
and $G_0$ replaced with $G_0'$ which is given by
\be
G_0' \equiv  m_\rho g_X Q_{E'} \left( {g_2 m_{E'} \over 2 {M_W} \sin\beta} \right)^2 
(A_{E'} \sin\alpha + \mu \cos\alpha) (A_{E'} \cos\alpha - \mu \sin\alpha)\  .
\ee

The relic density analysis is similar to that of \cite{Babu:2014pxa}. With the following variables 
\be
z \equiv {m_{H^0} \over T}, 
f_\rho \equiv {n_\rho \over T^3}, 
f_{H^0} \equiv {n_{H^0} \over T^3}, 
K \equiv {1.66 \sqrt{g_*} \over m_\text{Pl}}\  , 
\ee
{In the above $m_{H^0}$ is the mass of the heavy neutral Higgs,  $n_{H^0}$ is its 
number density, $n_\rho$ is the number density of $\rho$, and $g_*$ is the entropy 
degrees of freedom.  Further, } 
$T$ is the temperature of the  thermal bath, $H=KT^2$ is the Hubble constant, 
$ m_\text{Pl} = 1.22 \times 10^{19}$ GeV is the Planck mass, we obtain  {neglecting the
back reaction}\footnote{{The neglect of the back reaction is justified since the number density
of $\rho$ is small.} }
\be
{df_\rho (z) \over dz} = {\langle \Gamma(H^0\to h^0+\rho) \rangle \over K {m}_{H^0}^2}  z f_{H^0}(z)\  . 
\ee

The thermal average on the decay width is 
$\langle \Gamma\rangle = \Gamma {K_1(z)/ K_2(z)}$ 
where $K_{1,2}(z)$ are the modified Bessel functions, 
and $f_{H^0}(z) = z^2  K_2(z)/ (2 \pi^2)$ (assuming Maxwell-Boltzmann statistics), so we have 
\be
f_\rho(z) = {\Gamma(H^0\to h^0+\rho) \over K {m}_{H^0}^2} \int_{z'=z_0}^z dz' {K_1(z')} {z'^3 \over 2 \pi^2} \ ,
\ee
where $z_0 = {m}_{H^0}/T_\text{EW}$ with $T_\text{EW}$ being the electroweak phase transition temperature, 
and we neglected the temperature dependence of $K$.  
For the case where ${m}_{H^0}=500$ GeV and $T_\text{EW}=300$ GeV, 
the above integral gives an asymptotic value $\sim 0.2$ for $z >10$\  .
Thus we obtain (for ${m}_{H^0}=500$ GeV)
\be
f_\rho(z>10) \simeq {0.2 \times \Gamma(H^0\to h^0+\rho) \over K {m}_{H^0}^2} \  .
\ee
The quantity $n_\rho/s$ is conserved after $H^0$ disappears in the plasma, where $s$ is the entropy density. 
The current dark matter number density is then given by
\be
n_\rho^0 = (T_0)^3 {g_{*s}^0 \over g_{*s}^\text{freeze-out}} f_\rho(z=20) =  
(T_0)^3 {g_{*s}^0 \over g_{*s}^\text{freeze-out}} {0.2\times \Gamma(H^0\to h^0+\rho) \over K {m}_{H^0}^2}\  , 
\label{eq:rd}
\ee 
where $T_0=2.73$ K, $g_{*s}^0=3.91$ is the current degree of freedom, 
and $g_{*s}^\text{freeze-out}$ is the 
degree of freedom during freeze out so that 
$g_{*s}^\text{freeze-out} = g_{*s}(T\simeq m_{H^0}/20)$. 
The relic density due to $H^0$ decay is thus given by 
\be
\Omega_\rho h^2 = 
\frac{n_\rho^0 m_\rho}{\rho_c} h^2 \simeq
 {n_\rho^0 m_\rho \over 8 \times 10^{-47} ~\text{GeV}^{-4}}\  .
\ee

\subsection{{$\tilde{E}_1+\bar{\tilde{E}}_1  \to \rho + {h^0}$}}

\begin{figure}[t]
\begin{center}
\includegraphics[width=0.4\columnwidth]{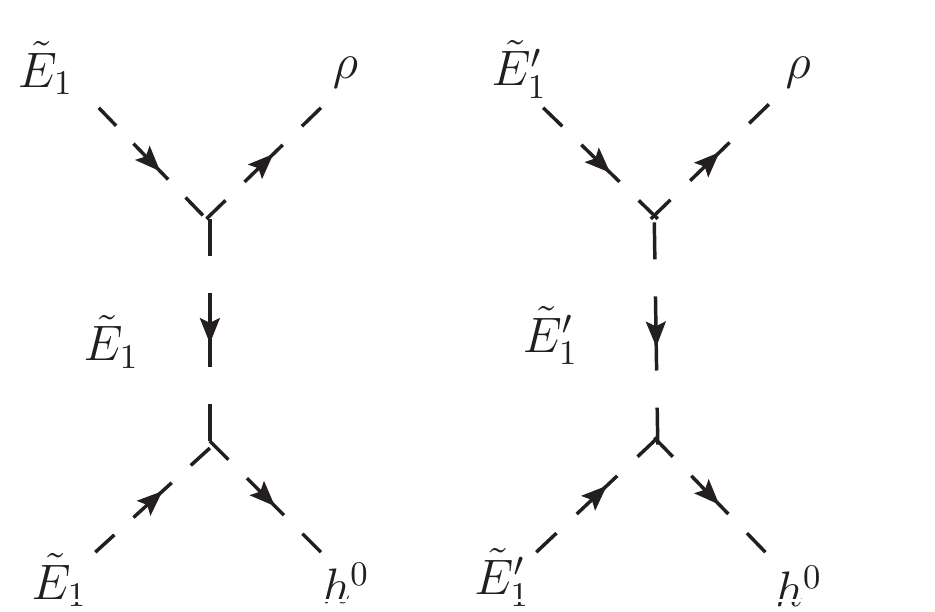}
\caption{Feynman {diagrams} for the process {$\tilde{E}_1 (\tilde{E}_1')+ \bar{\tilde{E}}_1  (\bar{\tilde{E}}_1')
\to \rho + h^0$}. 
}
\label{fig:feynRD2}
\end{center}
\end{figure}

The dark matter can also be generated via scalar annihilations. 
We first discuss $\tilde{E}$ scalar annihilation.
Among the two possible final states $\rho + h^0$ and $\rho+ \rho$, the 
$\rho+ \rho$ final state is suppressed by a factor of $m_\rho^4$.
Thus we  compute the matrix element for {$\tilde{E}_1+ \bar{\tilde{E}}_1 \to \rho + {h^0}$}
as shown in Fig.\ (\ref{fig:feynRD2}).   
Here we find 
\be
(v_\text{rel}\sigma) = {A \over s(s-m^2_{h^0})}\  , 
\ee
where 
\be
A\equiv \left[ m_\rho g_X Q_E {g_2 m_E \over 4 {M_W} \cos\beta} (-A_E \sin\alpha + \mu \cos\alpha) \sin(4\xi) \right]^2 
{1\over 8\pi m_{\tilde{E}_1}^2 } \  . 
\ee

The DM number density due to the annihilation process is 
given via the following Boltzmann equation (neglecting the initial DM density) 
\be
\dot{n}_\rho + 3 H n_\rho = n_{\tilde{E}_1}^2 \langle v_{\text{rel}} \sigma \rangle\  .
\ee
Defining the  variables $z \equiv {m_{\tilde{E}_1} / T}, 
f_{\tilde{E}_1} \equiv {n_{\tilde{E}_1} / T^3}$ (see \cite{Babu:2014pxa}) 
we obtain 
\be
{d f_\rho \over d z} = { m_{\tilde{E}_1} f_{\tilde{E}_1}^2  \over K z^2 } \langle v_{\text{rel}} \sigma \rangle\  .
\ee
The thermal average of the annihilation is given by  
\be
\langle v_{\text{rel}} \sigma \rangle = {1\over 16 T m_{\tilde{E}_1}^4 K_2^2(z)}
\int_{4m_{\tilde{E}_1}^2}^{\infty} 
s\sqrt{s-4m_{\tilde{E}_1}^2} K_1(\sqrt{s}/T) (v_{\text{rel}} \sigma) ds  \ . 
\ee
Thus we obtain 
\begin{align}
f_\rho &= {A \over 64 \pi^4 m_{\tilde{E}_1}^4 K} \int_1^{20} dz\ z^3 \int_{4 m_{\tilde{E}_1}}^{\infty} 
ds\ {\sqrt{s-4m_{\tilde{E}_1}^2} \over s-m_h^2} K_1(\sqrt{s}z/m_{\tilde{E}_1}) \nonumber\\
&\equiv {A \over 64 \pi^4 m_{\tilde{E}_1}^4 K} F(m_{\tilde{E}_1})\  ,
\end{align}
where we have used $f_{\tilde{E}_1} = z^2 K_2(z)/(2\pi^2)$
\footnote{For scalar mass higher than the electroweak phase transition temperature 
$T_\text{EW}=300$ GeV, the integral should start at $z={m_{\tilde{E}_1}/ T_\text{EW}}$. 
In this case using $z=1$ for the lower end of the integral overestimates the contribution 
due to scalar annihilations. However, as shown later, the scalar annihilation is 
subdominant 
to the $H^0$ decay process for the parameter space of interest
and is negligible in any  case for the parameter space of interest.}.
The 2-D integral only 
depends on the scalar mass $m_{\tilde{E}_1}$, and we found that 
the relation $F(m_{\tilde{E}_1}) \simeq 0.115 m_{\tilde{E}_1}$ can nicely 
approximate the integral for a large mass range, $m_{\tilde{E}_1}\in(10^2,10^7)$ GeV. 
The relic density calculation is similar to the decay process 
as discussed in Section \ref{sec:rddecay},  
once $f_\rho$ is known 
at ${\tilde E_1}$ freeze-out. 
Thus we obtain  the relic density due to annihilation 
\begin{align}
(\Omega_\rho h^2)_\text{ann} &= { m_\rho \over 8 \times 10^{-47} ~\text{GeV}^{-4}}
(T^0)^3 {g_{*s}^0 \over g_{*s}^\text{freeze-out}}
{A \over 64 \pi^4 m_{\tilde{E}_1}^4 K} 0.115 m_{\tilde{E}_1} \nonumber\\
&= 6 \times 10^{14} ~\text{GeV} {A \over m_{\tilde{E}_1}^3}\  . 
\end{align}
Similarly, one can compute the $\rho$ relic density due to the $\tilde{E}_1^\prime$ annihilations. 
For the $\tilde{E}_1^\prime$ annihilations, the quantity 
($A/m_{\tilde{E}_1}^3$) here has to be replaced by
($A^\prime / m_{\tilde{E}_1^\prime}^3$) where $A^\prime$ is given by 
\be
A^\prime 
\equiv \left[ m_\rho g_X Q_{E'} {g_2 m_{E'} \over 4 {M_W}  \sin\beta } 
{(A_{E'} \cos\alpha - \mu \sin\alpha)} \sin(4\xi') \right]^2 
{1\over 8\pi m_{\tilde{E}_1^\prime}^2 } \  .
\ee

\section{Numerical Analysis\label{sec5}}
 To fit the experimental data \cref{1} we need to compute the lifetime and the relic density of the $\rho$.
 This is done by  carrying out a scan in the parameter space where
 we take the ranges of the soft masses $M_1, M_2$, of the vectorlike masses $M_L, M_E$, of 
 the trilinear couplings $A_E, A_{E'}$ and of the fermion masses $m_E, m_{E'}$   generated by Yukawa couplings
  in the following ranges 
\begin{gather}
(M_1, M_2, M_L, M_E, A_E, A_{E'}, \mu) \in (10^2,10^5)~\GeV,  \\
{m_E  \in (100,246)~\GeV, m_{E'}  \in (100,300)~\GeV \ .}
\end{gather}
{
where 
 $m_E$ and $m_{E'}$  are defined so that  
  \begin{gather}
  m_E\equiv\frac{1}{\sqrt 2}  y v_d,  ~~ m_{E'} \equiv \frac{1}{\sqrt 2} y' v_u \ .
  \end{gather}
}
{Further we require that $y<2$ and $y'<2$, which put constraints on the $\tan\beta$ value such that 
$\pi/4<\beta<\arccos( m_E / (\sqrt{2} v))$. We also} 
take $g_X=1$, ${m}_{H^0}=500$ GeV, ${m}_{h^0}=125$ GeV,  
${m}_\rho=7$ keV, and $\alpha = \beta - \pi/2$ (see e.g.,\cite{Grinstein:2013npa}).
We investigate the possibility that
the dark matter constituted by $\rho$ contributes only a fraction of the relic density
for dark matter measured by WMAP which is $\Omega_{\rm WMAP} h^2 \simeq 0.11$~\cite{Komatsu:2010fb}.
Thus a desirable range is 
$R\equiv \Omega_\rho/\Omega_\text{DM} \in (0.01,0.1)$.
This would leave the other major component to be the neutralino for which the dark matter searches
 can be pursued 
in the direct and the indirect detection experiments. 
In the analysis of the relic density of $\rho$ we include all the allowed processes.
As discussed in \cref{sec4} and in  
{\cref{sec8}} the following processes contribute to the relic density

\begin{figure}[t]
\begin{center}
\includegraphics[width=0.6\columnwidth]{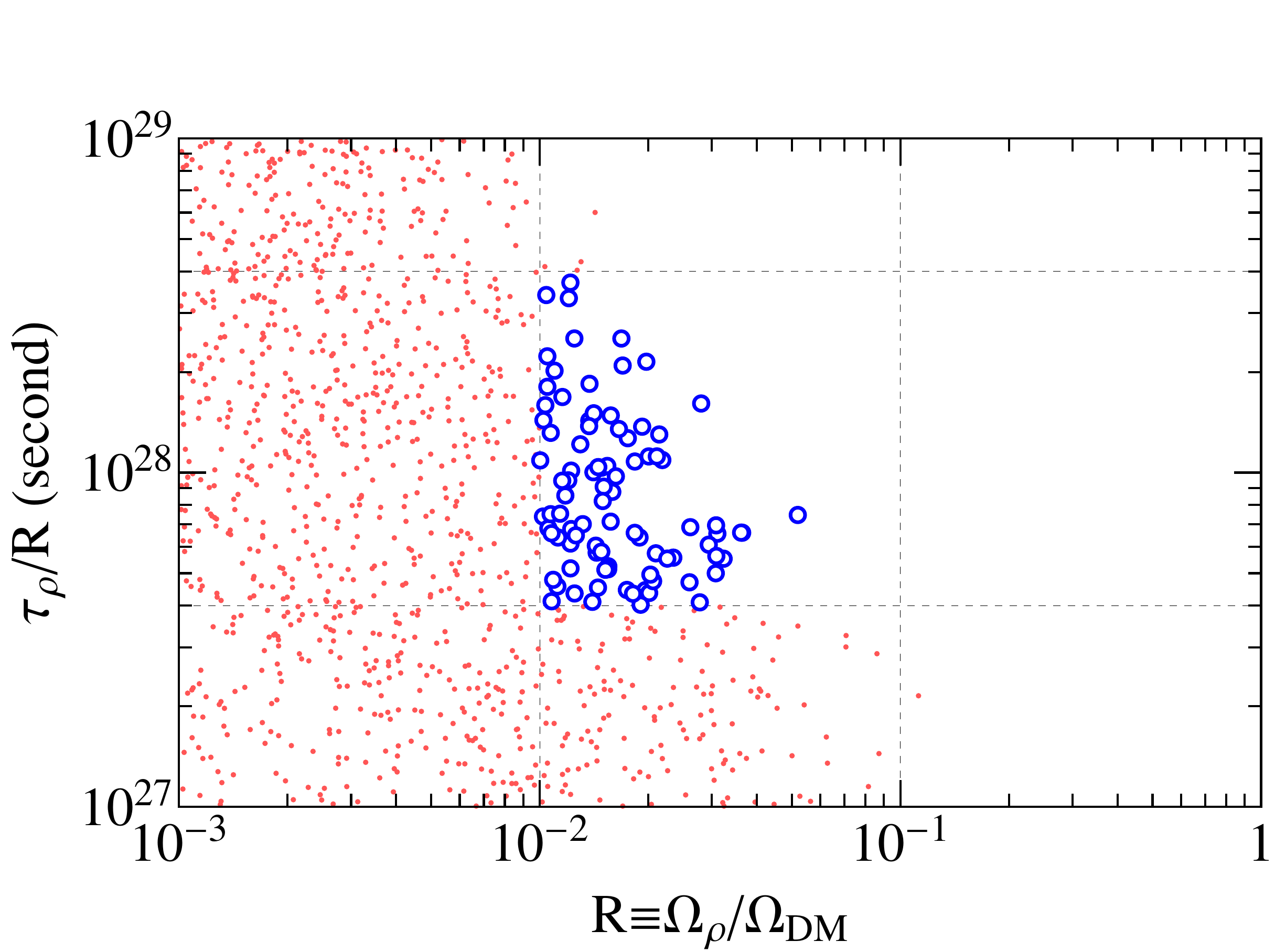}
\caption{\underline{Blue circles}:  {90} models out of the $2\times 10^5$ random scans in the parameter space 
satisfying the constraints
 $R\equiv \Omega_\rho / \Omega_\text{DM} \in (0.01,0.1)$ and $\tau_\rho/R \in 4 \times (10^{27}, 10^{28})$ s. 
\underline{Red points}: generic model points in the parameter space. 
}
\label{fig:scan}
\end{center}
\end{figure}

\begin{gather} 
H^0\to h^0+\rho\  , \\
\tilde E_1+\bar{ \tilde E}_1 \to \rho + h^0\   ,\\
\tilde E_1^{\prime}+ \bar{\tilde E}_1^{\prime } \to \rho + h^0\  .
\end{gather}
In addition there are other processes which are highly suppressed such as decays and annihilations 
with 2$\rho$ final states. We have computed  the processes listed above. Of these the one which
 produces the dominant component of the $\rho$ relic density is the process
 $H^0\to h^0 +\rho$ while the other processes are  subdominant. Thus the annihilation processes involving
  {$\tilde{E}_1+\bar{\tilde{E}}_1$} and {$\tilde{E}_1^\prime+\bar{\tilde{E}}_1^{\prime }$} together contribute  
at most 1\% of the number density of $\rho$, 
in the parameter space of interest.
An analysis of the $\rho$ lifetime vs $R$ is given in 
\cref{fig:scan}. Here one finds that there exist parameter choices  which fit the data on the 3.5 keV  emission line
 with $\rho$ being a subdominant component of the total relic density. Thus with $\rho$ constituting
 (1-10)\% of the relic density or even less it is possible to satisfy the constraint of \cref{1}.  
In \cref{tab:models} we exhibit a set of parameter points satisfying the relic density constraint 
discussed above and satisfying the constraint of \cref{1}. {It is interesting to ask what the impact 
is of this two component dark matter model with (1-10)\% of dark matter  constituted of $\rho$ 
and the rest made up of neutralinos on the direct detection of dark matter. The main impact is on the 
 comparison of theory with experiment.
 Thus one compares the theoretical value $r \sigma_{\chi_1^0 p}^{SI}$ 
where $r= \Omega_{\chi_1^0} h_0^2/ \Omega_{DM} h_0^2$ with experiment. For $r=0.9-0.99$, 
the theoretical predictions will be smaller by a factor of 0.9-0.99 which means that some of the parameter points that were 
eliminated by the current upper limits are still viable. So effectively inclusion of a second dark matter component
 increases the allowed 
{parameter}
 space of SUSY models and thus relaxes the dark matter constraints 
 from the direct detection of dark matter on these models. }

We note in passing that the 
Stueckelberg model will have  a $Z'$ which if stable will contribute to the dark matter density like the $\rho$.
{ However, unlike the $\rho$ it cannot decay into two photons. Further, is has no couplings to the standard model
particles. To circumvent this problem and allow $Z'$ to decay we can generate a small mixing between
$U(1)_X$ and a}  gauged $L_\mu-L_\tau$ (see, e.g., \cite{Feng:2012jn})
allowing the
$Z'$ to have a tiny coupling to the muon and to the tau neutrinos which permits the decays $Z'\to \nu_\mu \bar \nu_{\mu},
\nu_\tau\bar \nu_\tau$. On the other hand $\rho$ cannot couple to fermions and thus such decays 
are not allowed for the $\rho$ and the $\rho$ can only have photonic decays.
{We also note that after $Z'-Z''$ mixing where $Z''$ is the gauge boson of $L_\mu-L_\tau$
which also acquires a mass via the Stueckelberg mechanism, the  
 new leptons can annihilate to the standard model particles via the  $Z', Z''$ poles. Thus specifically
we will have 
{annihilation} processes such as {$E'\bar E'\to Z''\to \nu_\mu \bar \nu_{\mu},
\nu_\tau\bar \nu_\tau, \mu\bar \mu, \tau\bar \tau$  
 which  deplete the matter density of the new
leptons by resonant annihilation if  the mass of the $Z''$ is chosen to be in the vicinity of twice 
the mass of the new leptons. The analysis is similar to the one given in ~\cite{Feldman:2007wj}.}
Further, we give soft  masses to the 
  $U(1)_X$ and $U(1)_{L_\mu- L_\tau}$ gauginos which are large enough so they can decay into 
  the MSSM fields. Thus for $U(1)_X$  the coupling $\bar E' \lambda \tilde E'$ would decay $\lambda$ into $E'$ and $\tilde E'$ which in turn will annihilate to  the MSSM particles as discussed above.}\\

\begin{table}[htbp]
\begin{center}
\begin{tabular}{|c|c|c|c|c|c|c|c|c|c|c|}
\hline
Model & $M_1$ & $M_2$ & $M_L$ & $M_E$ & $m_E$ & $m_{E'}$ & $A_E$ & $A_{E'}$ & 
$\mu$ & $\tan\beta$ 
\\\hline
A& 4.21e3 & 2.55e3 & 3.05e2 & 2.02e2 & 1.73e2 & 1.59e2 & 4.32e4 & 1.06e2 & 6.06e4 & 1.47 \\\hline
B&  6.24e3 & 3.53e3 & 2.43e3 & 1.72e2 & 1.98e2 & 1.57e2 & 6.44e4 & 9.81e3 & 7.73e4 & 1.00 \\\hline
C&  3.17e4 & 5.07e3 & 8.43e2 & 3.72e3 & 1.03e2 & 2.79e2 & 1.01e4 & 8.88e4 & 7.07e2 & 1.61 \\\hline
D&  6.21e3 & 4.65e3 & 1.48e3 & 4.24e3 & 2.19e2 & 2.33e2 & 3.59e3 & 8.54e4 & 1.15e4 & 1.06 \\\hline
E&  4.88e4 & 2.65e3 & 1.05e3 & 1.41e3 & 2.06e2 & 1.41e2 & 1.15e2 & 4.62e2 & 6.45e4 & 1.17 \\\hline
\end{tabular}
\begin{tabular}{|c|c|c|c|c|c|c|c|c|}
\hline
Model & ${m}_{\tilde{E}_1}$ & ${m}_{\tilde{E}_2}$ & ${m}_{\tilde{E}_1'}$ & ${m}_{\tilde{E}_2'}$ 
& {$\Omega_\rho^{H^0}/\Omega_\text{DM}$} 
& {$\Omega_\rho^{\tilde{E}_1}/\Omega_\text{DM}$} 
& {$\Omega_\rho^{\tilde{E}_1'}/\Omega_\text{DM}$} 
& $\tau_\rho/R$ (s)  
\\\hline
A& 3.2e3 & 5.1e3 & 2.9e2 & 3.6e3 & 2.0e{-2} & 3.0e{-9} & 6.0e{-5} & 4.4e{27} \\\hline
B& 6.2e3 & 6.8e3 & 2.1e3 & 5.1e3 & 1.2e{-2} & 3.6e{-6} & 4.8e{-5} & 6.1e{27} \\\hline
C& 3.2e4 & 3.2e4 & 2.8e3 & 7.7e3 & 3.6e{-2} & 1.3e{-13} & 4.0e{-5} & 6.6e{27} \\\hline
D& 6.4e3 & 7.5e3 & 3.6e3 & 7.1e3 & 1.4e{-2} & 9.5e{-9} & 2.0e{-5} & 1.5e{28} \\\hline
E& 4.9e4 & 4.9e4 & 9.2e2 & 4.0e3 & 1.1e{-2} & 1.2e{-13} & 4.9e{-5} & 4.8e{27} \\\hline
\end{tabular}
\caption{Candidate models: A, B, C, D, and E. All the masses are in GeV.  
The quantities $\Omega_\rho^{H^0}$, 
$\Omega_\rho^{\tilde{E}_1}$, and 
$\Omega_\rho^{\tilde{E}_1'}$ 
are the contributions to the $\rho$ relic density due to 
the decay process $H^0 \to h^0 + \rho$, and the annihilation processes 
$\tilde{E}_1 +\bar{ \tilde{E}}_1 \to \rho + h^0 $ and 
$\tilde E'_1+ \bar{\tilde E}'_1 \to \rho + h^0$. 
$\Omega_\text{DM}$ is the total dark matter density which is taken to be 
$\Omega_\text{DM}h^2=0.11$ and $\tau_\rho$ is the dark matter lifetime. $R$ is the 
ratio {between} the {relic} density of the $\rho$ dark matter and the total dark matter. 
 }
\label{tab:models}
\end{center}
\end{table}

\section{\label{sec6}Conclusion}
In this work we have given an analysis of the 3.5 keV emission  line emanating from galaxy clusters 
as seen by two experiments.  A possible explanation of the monochromatic nature of the radiation
is that it originates from the decay of a 7 keV particle. In this work we identify this particle as 
a scalar  $\rho$  
that appears in a supersymmetric $U(1)_X$ \st of models with the standard model gauge group.
In such an extension, the $\rho$ couples only to scalar fields that carry a $U(1)_X$ quantum number. 
The proposed $U(1)_X$ extension contains vectorlike multiplets which are charged under $SU(2)_L\times U(1)_Y$ 
as well as under $U(1)_X$.  Thus the scalars of the vectorlike multiplet 
 couple to $\rho$ as well as to the Higgs field and to the photon. These vectorlike couplings  allow
 the decay of the $\rho$ to two photons via triangle loops involving scalar particles. An important constraint
 on the lifetime of the $\rho$ arises from the fraction that $\rho$ contributes to the dark matter relic density. 
 The relic density of the $\rho$  arises only after electroweak symmetry breaking.
 Thus below the electroweak symmetry breaking scale  the CP even Higgses can decay to a $\rho$
 and there are various   decay channels such as $h^0\to \rho + \rho$,
 $H^0\to \rho + h^0$, $H^0\to \rho +\rho$. Additionally annihilation can contribute
 to the relic density such as via the process  {$\tilde E_1 + \bar{\tilde E}_1\to h^0 +\rho$} and the process
  {$\tilde E_1^\prime + \bar{\tilde E}_1^{\prime } \to h^0 +\rho$}. However, the dominant process that contributes to the relic 
  density turns out to be $H^0\to \rho+ h^0$.  
A simultaneous analysis of the relic density and of the $\rho$ lifetime is needed to fit the data.
   In the analysis presented in this work we are able to fit the data with $\rho$ as a subdominant
component of dark matter. Thus we have a multicomponent dark matter 
model where the emission line arises from a 7 keV scalar particle while the rest is constituted of neutralino
dark matter which can be detected in direct detection experiments such as XENON1T\cite{Aprile:2012zx},
 SuperCDMS\cite{Cabrera:2005zz}  and LUX~\cite{Akerib:2013tjd}. 
 Finally we note that  our mechanism for generating a 3.5 keV line as well as the implications of the model are very different 
 from other models that have recently been proposed
 \cite{ Cline:2014eaa,Babu:2014pxa,Conlon:2014xsa,Robinson:2014bma,Lee:2014koa,Okada:2014zea,Modak:2014vva,Dudas:2014ixa,Queiroz:2014yna,Demidov:2014hka, Ko:2014xda, Nakayama:2014cza, Bomark:2014yja,Liew:2014gia,Allahverdi:2014dqa, Kolda:2014ppa,Bezrukov:2014nza,Cicoli:2014bfa, Baek:2014qwa,Choi:2014tva,Nakayama:2014ova,Frandsen:2014lfa,Kong:2014gea,Aisati:2014nda, 
 Krall:2014dba,Abazajian:2014gza, Jaeckel:2014qea,Higaki:2014zua, Finkbeiner:2014sja,Ishida:2014dlp,Boyarsky:2014jta,Baek:2014poa,Nakayama:2014rra,Chakraborty:2014tma}. \\

\noindent
{\em Acknowledgments}:
This research was supported in part by the NSF Grant PHY-1314774, NSF of China (under Grants 11275101, 11135003),
XSEDE-TG-PHY110015, and NERSC-DE-AC02-05CH1123.

\appendix

\section{Interactions of the vectorlike multiplet}
\label{sec7}

 We discuss now the interactions that are needed for the analysis in this work
 using the results of \cite{Ibrahim:2008gg,Kors:2004ri,Gunion:1984yn}.  The photonic interactions are given by 
 \begin{gather}
  {\cal L}_{\gamma}
=i q_E e (\tilde E_L^*\stackrel{\leftrightarrow}{\partial_\mu}\tilde E_L 
 +\tilde E_R^*\stackrel{\leftrightarrow}{\partial_\mu}\tilde E_R) A^\mu
+ i q_{E'} e (\tilde E_L^{{\prime *}}\stackrel{\leftrightarrow}{\partial_\mu}\tilde E_L' 
 +\tilde E_R^{{\prime *}}\stackrel{\leftrightarrow}{\partial_\mu}\tilde E_R' ) A^\mu \  , 
 \end{gather}
 where $q_E= q_{E'}=1$.
 The interactions of $h^0/H^0$ with the charged particles of the  vector multiplet are given by
 (see, e,g,  ~\cite{Gunion:1984yn}, Eq. (4.19)) 
 \begin{gather}
 {\cal L}_{EEh^0/H^0}= - \frac{g_2 m_{E}}{2 M_W \cos\beta} ( \tilde E_R^{*} \tilde E_L + \tilde E_L^{*} \tilde E_R) 
 \left[(A_{E}  \cos\alpha + \mu \sin\alpha) H^0  + 
 (-A_{E}  \sin\alpha + \mu \cos\alpha) h^0\right]  ,  \\
 {\cal L}_{E'E'h^0/H^0}= - \frac{g_2 m_{E'}}{2 M_W \sin\beta} ( \tilde E_R^{{\prime *}} \tilde E_L' + \tilde E_L^{{\prime *}} \tilde E_R') 
 \left[(A_{E'}  \sin\alpha + \mu \cos\alpha) H^0  + 
 (A_{E'}  \cos\alpha - \mu \sin\alpha) h^0\right]\  , 
\end{gather}
where $\alpha$ is the Higgs mixing angle defined by 
 \begin{gather} 
H_d^1 = \frac{1}{\sqrt 2} \left[ v_d + H^0 \cos\alpha -h^0 \sin\alpha\right] \ ,  \\
H_u^2  = \frac{1}{\sqrt 2} \left[ v_u + H^0 \sin\alpha +h^0 \cos\alpha\right] \  .
 \end{gather}
The $\rho$ couplings from the \st are given by  \cite{Kors:2004ri}
\begin{align}
{\cal L}_{st} & =m_\rho\rho\sum_i g_X Q_i \bar z_i z_i \\
&= m_\rho g_X \rho \left[ Q_E \tilde E_L^* \tilde E_L + Q_{E^c} \tilde E_R^* \tilde E_R
+ Q_{E'} \tilde E_L^{{\prime *}} \tilde E_L' + Q_{E^{{\prime c}}} \tilde E_R^{{\prime *}} \tilde E_R'
\right]\   , 
\end{align}
where $Q_E= -Q_{E^c}= Q_{E'} = - Q_{E^{{\prime c}}}$.
Next we write the above interactions in a mass diagonal basis using the following relations  
\begin{gather} 
( \tilde E_R^* \tilde E_L + \tilde E_L^* \tilde E_R) 
=\sin 2 \xi  ( -\tilde E_1^* \tilde E_1 + \tilde E_2^* \tilde E_2) 
+\cos 2 \xi  (\tilde E_1^* \tilde E_2 + \tilde E_2^* \tilde E_1)\  , \\
( \tilde E_L^* \tilde E_L - \tilde E_R^* \tilde E_R) 
=\cos 2 \xi  ( \tilde E_1^* \tilde E_1 -\tilde E_2^* \tilde E_2) 
+\sin 2 \xi  (\tilde E_1^* \tilde E_2 + \tilde E_2^* \tilde E_1).
\end{gather}
{The relations above hold in the approximation when the off-diagonal terms in 
\cref{slep1} are neglected.}
The relevant interactions in the mass diagonal  basis are given by 
\begin{gather}
 {\cal L}= {\cal L}_{st}+  {\cal L}_{\gamma}+ {\cal L}_{EEh^0/H^0}\  ,\\
\label{ff1}
{\cal L}_{st}= m_\rho g_X Q_E \rho
\left[
\cos 2 \xi  ( \tilde E_1^* \tilde E_1 -\tilde E_2^* \tilde E_2) 
+\sin 2 \xi  (\tilde E_1^* \tilde E_2 + \tilde E_2^* \tilde E_1)\right]\nonumber\\
+ m_\rho g_X Q_{E'} \rho
\left[
\cos 2 \xi'  ( \tilde E_1^{{\prime *} }\tilde E_1' -\tilde E_2^{{\prime *}} \tilde E_2') 
+\sin 2 \xi'  (\tilde E_1^{{\prime *}} \tilde E_2' + \tilde E_2^{{\prime *}} \tilde E_1')\right] \  ,\\
\label{ff2}
{\cal L}_{\gamma}
=i  e q_E (\tilde E_1^*\stackrel{\leftrightarrow}{\partial_\mu}\tilde E_1 
 +\tilde E_2^*\stackrel{\leftrightarrow}{\partial_\mu}\tilde E_2 ) A^\mu\nonumber\\
+ i  e q_{E'} (\tilde E_1^{{\prime *}}\stackrel{\leftrightarrow}{\partial_\mu}\tilde E_1' 
 +\tilde E_2^{{\prime *}}\stackrel{\leftrightarrow}{\partial_\mu}\tilde E_2' ) A^\mu\  , \\
 \label{ff3}  
{\cal L}_{EEh^0/H^0}= - \frac{g_2 m_E}{2 M_W \cos\beta}
 \left[(A_E  \cos\alpha + \mu \sin\alpha) H^0  + 
 (-A_E  \sin\alpha + \mu \cos\alpha) h^0\right]\nonumber\\
 \times 
\left[\sin 2 \xi  ( -\tilde E_1^* \tilde E_1 + \tilde E_2^* \tilde E_2) 
+\cos 2 \xi  (\tilde E_1^* \tilde E_2 + \tilde E_2^* \tilde E_1)\right]\nonumber\\
- \frac{g_2 m_{E'}}{2 M_W \sin\beta}
 \left[(A_{E'}  \sin\alpha + \mu \cos\alpha) H^0  + 
 (A_{E'}  \cos\alpha - \mu \sin\alpha) h^0\right]\nonumber\\
 \times 
\left[\sin 2 \xi'  ( -\tilde E_1^{{\prime *}} \tilde E_1' + \tilde E_2^{{\prime *}} \tilde E_2') 
+\cos 2 \xi'  (\tilde E_1^{{\prime *}} \tilde E_2' + \tilde E_2^{{\prime *}} \tilde E_1')\right]\ .
 \label{ff4}
 \end{gather}
  These are the interactions that are used in the analysis of the paper.

\section{$h^0 h^0\to \rho\rho$}
\label{sec8}

The interaction between $\rho$ and the Higgs boson can be parameterized as follows 
\be
{\cal L} =m_\rho {g_{\rho h h} \over 2!} \rho {h^0} {h^0} \  . 
\ee
The averaged matrix element square for the process (${h^0} {h^0} \to \rho \rho$) is given by 
(exchanging a Higgs in both t- and u- channels)
\be
\overline{| {\cal M} |^2} = m_\rho^4 g_{\rho h h}^4 \left[  
{1\over (p_3-p_1)^2-m^2_{h^0}} + {1\over (p_4-p_1)^2-m^2_{h^0}}  
\right]^2 \  .
\ee 
The annihilation rate is given by 
\bea
v_\text{rel} \sigma  
={1\over 2} {\beta_f \over 16 \pi s} \int_{-1}^1 dz \overline{| {\cal M} |^2} 
=m_\rho^4  {2 g_{\rho h h}^4 \over \pi s^3} {\beta_f \over (1+\beta_f^2)^2} 
\left[ 
{a^2 \over a^2 -1} + {a\over 2} \ln
\left({a+1\over a-1}\right)
\right]\  . 
\eea
Here $\beta_i \equiv \sqrt{1-4m_h^2/s}$, $\beta_f \equiv \sqrt{1-4m_\rho^2/s}$, 
$a \equiv (1+\beta_f^2)/(2\beta_i \beta_f)$. 
This cross section is suppressed by 
$m_{\rho}^4/s^2 \simeq (m_\rho/2m_{h^0})^4 \sim 10^{-30}$.
In the limit $\beta_i \to 0$ and $\beta_f \to 1$, we 
get 
\be v_\text{rel} \sigma  = m_\rho^4  {g_{\rho h h}^4 \over \pi s^3} 
{\simeq 10^{-36}~\text{GeV}^{-2}}\  .
\label{175} 
\ee
For the annihilation process to make a significant contribution 
$(v_\text{rel} \sigma)$ should be the range $10^{-23}-10^{-24} ~\GeV^{-2}$. 
Thus  the result of \cref{175} is far too small.

%

\begin{thebibliography}{10}%
\makeatletter
\providecommand{\hrefCMSnoop }[0]{\@secondoftwo}%
\makeatother
\providecommand{\doi}{\texttt{doi:}\begingroup \urlstyle{tt}\Url}

\bibitem{Bulbul:2014sua}
\hrefCMSnoop {} {E.~Bulbul, M.~Markevitch, A.~Foster{ et~al.}, ``{Detection of
  An Unidentified Emission Line in the Stacked X-ray spectrum of Galaxy
  Clusters}'',}
\href{http://www.arXiv.org/abs/1402.2301}{\texttt{ arXiv:1402.2301}}.

\bibitem{Boyarsky:2014jta}
\hrefCMSnoop {} {A.~Boyarsky, O.~Ruchayskiy, D.~Iakubovskyi{ et~al.}, ``{An
  unidentified line in X-ray spectra of the Andromeda galaxy and Perseus galaxy
  cluster}'',}
\href{http://www.arXiv.org/abs/1402.4119}{\texttt{ arXiv:1402.4119}}.

\bibitem{Babu:2014pxa}
\hrefCMSnoop {} {K.~Babu and R.~N. Mohapatra, ``{7 keV Scalar Dark Matter and
  the Anomalous Galactic X-ray Spectrum}'',}
\href{http://www.arXiv.org/abs/1404.2220}{\texttt{ arXiv:1404.2220}}.


\bibitem{Frandsen:2014lfa}
\hrefCMSnoop {} {M.~T. Frandsen, F.~Sannino, I.~M. Shoemaker{ et~al.}, ``{X-ray
  Lines from Dark Matter: The Good, The Bad, and The Unlikely}'',}
\href{http://www.arXiv.org/abs/1403.1570}{\texttt{ arXiv:1403.1570}}.

\bibitem{Cline:2014eaa}
\hrefCMSnoop {} {J.~M. Cline, Y.~Farzan, Z.~Liu{ et~al.}, ``{3.5 keV X-rays as
  the "21 cm line" of dark atoms, and a link to light sterile neutrinos}'',}
\href{http://www.arXiv.org/abs/1404.3729}{\texttt{ arXiv:1404.3729}}.

\bibitem{Kors:2004dx}
\hrefCMSnoop {} {B.~Kors and P.~Nath, ``{A Stueckelberg extension of the
  standard model}'',} \textit{ Phys.Lett.} \textbf{ B586} (2004) 366--372,
  \href{http://dx.doi.org/10.1016/j.physletb.2004.02.051}{\doi{10.1016/j.physletb.2004.02.051}},
\href{http://www.arXiv.org/abs/hep-ph/0402047}{\texttt{ arXiv:hep-ph/0402047}}.

\bibitem{Kors:2004ri}
\hrefCMSnoop {} {B.~Kors and P.~Nath, ``{A Supersymmetric Stueckelberg U(1)
  extension of the MSSM}'',} \textit{ JHEP} \textbf{ 0412} (2004) 005,
  \href{http://dx.doi.org/10.1088/1126-6708/2004/12/005}{\doi{10.1088/1126-6708/2004/12/005}},
\href{http://www.arXiv.org/abs/hep-ph/0406167}{\texttt{ arXiv:hep-ph/0406167}}.

\bibitem{Kors:2004iz}
\hrefCMSnoop {} {B.~Kors and P.~Nath, ``{How Stueckelberg extends the standard
  model and the MSSM}'',}
\href{http://www.arXiv.org/abs/hep-ph/0411406}{\texttt{ arXiv:hep-ph/0411406}}.

\bibitem{Kors:2005uz}
\hrefCMSnoop {} {B.~Kors and P.~Nath, ``{Aspects of the Stueckelberg
  extension}'',} \textit{ JHEP} \textbf{ 0507} (2005) 069,
  \href{http://dx.doi.org/10.1088/1126-6708/2005/07/069}{\doi{10.1088/1126-6708/2005/07/069}},
\href{http://www.arXiv.org/abs/hep-ph/0503208}{\texttt{ arXiv:hep-ph/0503208}}.

\bibitem{Cheung:2007ut}
\hrefCMSnoop {} {K.~Cheung and T.-C. Yuan, ``{Hidden fermion as milli-charged
  dark matter in Stueckelberg Z- prime model}'',} \textit{ JHEP} \textbf{ 0703}
  (2007) 120,
  \href{http://dx.doi.org/10.1088/1126-6708/2007/03/120}{\doi{10.1088/1126-6708/2007/03/120}},
\href{http://www.arXiv.org/abs/hep-ph/0701107}{\texttt{ arXiv:hep-ph/0701107}}.

\bibitem{Feldman:2007nf}
\hrefCMSnoop {} {D.~Feldman, Z.~Liu, and P.~Nath, ``{The Stueckelberg extension
  and milli weak and milli charged dark matter}'',} \textit{ AIP Conf.Proc.}
  \textbf{ 939} (2007) 50--58,
  \href{http://dx.doi.org/10.1063/1.2803786}{\doi{10.1063/1.2803786}},
\href{http://www.arXiv.org/abs/0705.2924}{\texttt{ arXiv:0705.2924}}.

\bibitem{Feldman:2007wj}
\hrefCMSnoop {} {D.~Feldman, Z.~Liu, and P.~Nath, ``{The Stueckelberg Z-prime
  Extension with Kinetic Mixing and Milli-Charged Dark Matter From the Hidden
  Sector}'',} \textit{ Phys.Rev.} \textbf{ D75} (2007) 115001,
  \href{http://dx.doi.org/10.1103/PhysRevD.75.115001}{\doi{10.1103/PhysRevD.75.115001}},
\href{http://www.arXiv.org/abs/hep-ph/0702123}{\texttt{ arXiv:hep-ph/0702123}}.

\bibitem{Feldman:2006wb}
\hrefCMSnoop {} {D.~Feldman, Z.~Liu, and P.~Nath, ``{The Stueckelberg $Z$ Prime
  at the LHC: Discovery Potential, Signature Spaces and Model
  Discrimination}'',} \textit{ JHEP} \textbf{ 0611} (2006) 007,
  \href{http://dx.doi.org/10.1088/1126-6708/2006/11/007}{\doi{10.1088/1126-6708/2006/11/007}},
\href{http://www.arXiv.org/abs/hep-ph/0606294}{\texttt{ arXiv:hep-ph/0606294}}.

\bibitem{Liu:2011di}
\hrefCMSnoop {} {Z.~Liu, P.~Nath, and G.~Peim, ``{An Explanation of the CDF
  Dijet Anomaly within a $U(1)_X$ Stueckelberg Extension}'',} \textit{
  Phys.Lett.} \textbf{ B701} (2011) 601--604,
  \href{http://dx.doi.org/10.1016/j.physletb.2011.06.045}{\doi{10.1016/j.physletb.2011.06.045}},
\href{http://www.arXiv.org/abs/1105.4371}{\texttt{ arXiv:1105.4371}}.

\bibitem{Feldman:2011ms}
\hrefCMSnoop {} {D.~Feldman, P.~Fileviez~Perez, and P.~Nath, ``{R-parity
  Conservation via the Stueckelberg Mechanism: LHC and Dark Matter Signals}'',}
  \textit{ JHEP} \textbf{ 1201} (2012) 038,
  \href{http://dx.doi.org/10.1007/JHEP01(2012)038}{\doi{10.1007/JHEP01(2012)038}},
\href{http://www.arXiv.org/abs/1109.2901}{\texttt{ arXiv:1109.2901}}.

\bibitem{Feng:2012jn}
\hrefCMSnoop {} {W.-Z. Feng, P.~Nath, and G.~Peim, ``{Cosmic Coincidence and
  Asymmetric Dark Matter in a Stueckelberg Extension}'',} \textit{ Phys.Rev.}
  \textbf{ D85} (2012) 115016,
  \href{http://dx.doi.org/10.1103/PhysRevD.85.115016}{\doi{10.1103/PhysRevD.85.115016}},
\href{http://www.arXiv.org/abs/1204.5752}{\texttt{ arXiv:1204.5752}}.

\bibitem{Perez:2014gta}
\hrefCMSnoop {} {P.~Fileviez~Perez and S.~Spinner, ``{The Higgs Mass and the
  Stueckelberg Mechanism in Supersymmetry}'',} \textit{ Phys.Rev.} \textbf{
  D89} (2014) 095004,
  \href{http://dx.doi.org/10.1103/PhysRevD.89.095004}{\doi{10.1103/PhysRevD.89.095004}},
\href{http://www.arXiv.org/abs/1401.7327}{\texttt{ arXiv:1401.7327}}.

\bibitem{Santos:2014xka}
\hrefCMSnoop {} {A.~L.~d. Santos and D.~Hadjimichef, ``{Astrophysical aspects
  of milli-charged dark matter in a Higgs-Stueckelberg model}'',}
\href{http://www.arXiv.org/abs/1405.4282}{\texttt{ arXiv:1405.4282}}.

\bibitem{Feng:2013mea}
\hrefCMSnoop {} {W.-Z. Feng and P.~Nath, ``{Higgs diphoton rate and mass
  enhancement with vectorlike leptons and the scale of supersymmetry}'',}
  \textit{ Phys.Rev.} \textbf{ D87} (2013), no.~7, 075018,
  \href{http://dx.doi.org/10.1103/PhysRevD.87.075018}{\doi{10.1103/PhysRevD.87.075018}},
\href{http://www.arXiv.org/abs/1303.0289}{\texttt{ arXiv:1303.0289}}.

\bibitem{Grinstein:2013npa}
\hrefCMSnoop {} {B.~Grinstein and P.~Uttayarat, ``{Carving Out Parameter Space
  in Type-II Two Higgs Doublets Model}'',} \textit{ JHEP} \textbf{ 1306} (2013)
  094, \href{http://dx.doi.org/10.1007/JHEP09(2013)110,
  10.1007/JHEP06(2013)094}{\doi{10.1007/JHEP09(2013)110,
  10.1007/JHEP06(2013)094}},
\href{http://www.arXiv.org/abs/1304.0028}{\texttt{ arXiv:1304.0028}}.

\bibitem{Komatsu:2010fb}
\hrefCMSnoop {} {{ WMAP Collaboration} Collaboration, ``{Seven-Year Wilkinson
  Microwave Anisotropy Probe (WMAP) Observations: Cosmological
  Interpretation}'',} \textit{ Astrophys.J.Suppl.} \textbf{ 192} (2011) 18,
  \href{http://dx.doi.org/10.1088/0067-0049/192/2/18}{\doi{10.1088/0067-0049/192/2/18}},
\href{http://www.arXiv.org/abs/1001.4538}{\texttt{ arXiv:1001.4538}}.

\bibitem{Aprile:2012zx}
\hrefCMSnoop {} {{ XENON1T collaboration} Collaboration, ``{The XENON1T Dark
  Matter Search Experiment}'',}
\href{http://www.arXiv.org/abs/1206.6288}{\texttt{ arXiv:1206.6288}}.

\bibitem{Cabrera:2005zz}
\hrefCMSnoop {} {{ SuperCDMS} Collaboration, ``SuperCDMS Development
  Project'',} (April, 2005). Fermilab Program Advisory Committee.

\bibitem{Akerib:2013tjd}
\hrefCMSnoop {} {{ LUX Collaboration} Collaboration, ``{First results from the
  LUX dark matter experiment at the Sanford Underground Research Facility}'',}
\href{http://www.arXiv.org/abs/1310.8214}{\texttt{ arXiv:1310.8214}}.

\bibitem{Conlon:2014xsa}
\hrefCMSnoop {} {J.~P. Conlon and F.~V. Day, ``{3.55 keV photon lines from
  axion to photon conversion in the Milky Way and M31}'',}
\href{http://www.arXiv.org/abs/1404.7741}{\texttt{ arXiv:1404.7741}}.

\bibitem{Robinson:2014bma}
\hrefCMSnoop {} {D.~J. Robinson and Y.~Tsai, ``{A Dynamical Framework for KeV
  Dirac Neutrino Warm Dark Matter}'',}
\href{http://www.arXiv.org/abs/1404.7118}{\texttt{ arXiv:1404.7118}}.

\bibitem{Lee:2014koa}
\hrefCMSnoop {} {H.~M. Lee, ``{Magnetic dark matter for the X-ray line at 3.55
  keV}'',}
\href{http://www.arXiv.org/abs/1404.5446}{\texttt{ arXiv:1404.5446}}.

\bibitem{Okada:2014zea}
\hrefCMSnoop {} {H.~Okada and T.~Toma, ``{The 3.55 keV X-ray Line Signal from
  Excited Dark Matter in Radiative Neutrino Model}'',}
\href{http://www.arXiv.org/abs/1404.4795}{\texttt{ arXiv:1404.4795}}.

\bibitem{Modak:2014vva}
\hrefCMSnoop {} {K.~P. Modak, ``{3.5 keV X-ray Line Signal from Decay of
  Right-Handed Neutrino due to Transition Magnetic Moment}'',}
\href{http://www.arXiv.org/abs/1404.3676}{\texttt{ arXiv:1404.3676}}.

\bibitem{Dudas:2014ixa}
\hrefCMSnoop {} {E.~Dudas, L.~Heurtier, and Y.~Mambrini, ``{Generating X-ray
  lines from annihilating dark matter}'',}
\href{http://www.arXiv.org/abs/1404.1927}{\texttt{ arXiv:1404.1927}}.

\bibitem{Queiroz:2014yna}
\hrefCMSnoop {} {F.~S. Queiroz and K.~Sinha, ``{The Poker Face of the Majoron
  Dark Matter Model: LUX to keV Line}'',}
\href{http://www.arXiv.org/abs/1404.1400}{\texttt{ arXiv:1404.1400}}.

\bibitem{Demidov:2014hka}
\hrefCMSnoop {} {S.~Demidov and D.~Gorbunov, ``{SUSY in the sky or keV
  signature of sub-GeV gravitino dark matter}'',}
\href{http://www.arXiv.org/abs/1404.1339}{\texttt{ arXiv:1404.1339}}.

\bibitem{Ko:2014xda}
\hrefCMSnoop {} {P.~Ko, Z.~kang, T.~Li{ et~al.}, ``{Natural $X$-ray Lines from
  the Low Scale Supersymmetry Breaking}'',}
\href{http://www.arXiv.org/abs/1403.7742}{\texttt{ arXiv:1403.7742}}.

\bibitem{Nakayama:2014cza}
\hrefCMSnoop {} {K.~Nakayama, F.~Takahashi, and T.~T. Yanagida, ``{Anomaly-free
  flavor models for Nambu-Goldstone bosons and the 3.5 keV X-ray line
  signal}'',}
\href{http://www.arXiv.org/abs/1403.7390}{\texttt{ arXiv:1403.7390}}.

\bibitem{Bomark:2014yja}
\hrefCMSnoop {} {N.~E. Bomark and L.~Roszkowski, ``{The 3.5 keV X-ray line from
  decaying gravitino dark matter}'',}
\href{http://www.arXiv.org/abs/1403.6503}{\texttt{ arXiv:1403.6503}}.

\bibitem{Liew:2014gia}
\hrefCMSnoop {} {S.~P. Liew, ``{Axino dark matter in light of an anomalous
  X-ray line}'',}
\href{http://www.arXiv.org/abs/1403.6621}{\texttt{ arXiv:1403.6621}}.

\bibitem{Allahverdi:2014dqa}
\hrefCMSnoop {} {R.~Allahverdi, B.~Dutta, and Y.~Gao, ``{keV Photon Emission
  from Light Nonthermal Dark Matter}'',}
\href{http://www.arXiv.org/abs/1403.5717}{\texttt{ arXiv:1403.5717}}.

\bibitem{Kolda:2014ppa}
\hrefCMSnoop {} {C.~Kolda and J.~Unwin, ``{X-ray lines from R-parity violating
  decays of keV sparticles}'',}
\href{http://www.arXiv.org/abs/1403.5580}{\texttt{ arXiv:1403.5580}}.

\bibitem{Bezrukov:2014nza}
\hrefCMSnoop {} {F.~Bezrukov and D.~Gorbunov, ``{Relic Gravity Waves and 7 keV
  Dark Matter from a GeV scale inflaton}'',}
\href{http://www.arXiv.org/abs/1403.4638}{\texttt{ arXiv:1403.4638}}.

\bibitem{Cicoli:2014bfa}
\hrefCMSnoop {} {M.~Cicoli, J.~P. Conlon, M.~C.~D. Marsh{ et~al.}, ``{A 3.55
  keV Photon Line and its Morphology from a 3.55 keV ALP Line}'',}
\href{http://www.arXiv.org/abs/1403.2370}{\texttt{ arXiv:1403.2370}}.

\bibitem{Baek:2014qwa}
\hrefCMSnoop {} {S.~Baek and H.~Okada, ``{7 keV Dark Matter as X-ray Line
  Signal in Radiative Neutrino Model}'',}
\href{http://www.arXiv.org/abs/1403.1710}{\texttt{ arXiv:1403.1710}}.

\bibitem{Choi:2014tva}
\hrefCMSnoop {} {K.-Y. Choi and O.~Seto, ``{X-ray line signal from decaying
  axino warm dark matter}'',}
\href{http://www.arXiv.org/abs/1403.1782}{\texttt{ arXiv:1403.1782}}.

\bibitem{Nakayama:2014ova}
\hrefCMSnoop {} {K.~Nakayama, F.~Takahashi, and T.~T. Yanagida, ``{The 3.5 keV
  X-ray line signal from decaying moduli with low cutoff scale}'',}
\href{http://www.arXiv.org/abs/1403.1733}{\texttt{ arXiv:1403.1733}}.


\bibitem{Kong:2014gea}
\hrefCMSnoop {} {K.~Kong, J.-C. Park, and S.~C. Park, ``{X-ray line signal from
  7 keV axino dark matter decay}'',}
\href{http://www.arXiv.org/abs/1403.1536}{\texttt{ arXiv:1403.1536}}.

\bibitem{Aisati:2014nda}
\hrefCMSnoop {} {C.~E. Aisati, T.~Hambye, and T.~Scarna, ``{Can a millicharged
  dark matter particle emit an observable gamma-ray line?}'',}
\href{http://www.arXiv.org/abs/1403.1280}{\texttt{ arXiv:1403.1280}}.

\bibitem{Krall:2014dba}
\hrefCMSnoop {} {R.~Krall, M.~Reece, and T.~Roxlo, ``{Effective field theory
  and keV lines from dark matter}'',}
\href{http://www.arXiv.org/abs/1403.1240}{\texttt{ arXiv:1403.1240}}.

\bibitem{Abazajian:2014gza}
\hrefCMSnoop {} {K.~N. Abazajian, ``{Resonantly-Produced 7 keV Sterile Neutrino
  Dark Matter Models and the Properties of Milky Way Satellites}'',} \textit{
  Phys.Rev.Lett.} \textbf{ 112} (2014) 161303,
  \href{http://dx.doi.org/10.1103/PhysRevLett.112.161303}{\doi{10.1103/PhysRevLett.112.161303}},
\href{http://www.arXiv.org/abs/1403.0954}{\texttt{ arXiv:1403.0954}}.

\bibitem{Jaeckel:2014qea}
\hrefCMSnoop {} {J.~Jaeckel, J.~Redondo, and A.~Ringwald, ``{A 3.55 keV hint
  for decaying axion-like particle dark matter}'',}
\href{http://www.arXiv.org/abs/1402.7335}{\texttt{ arXiv:1402.7335}}.

\bibitem{Higaki:2014zua}
\hrefCMSnoop {} {T.~Higaki, K.~S. Jeong, and F.~Takahashi, ``{The 7 keV axion
  dark matter and the X-ray line signal}'',} \textit{ Phys.Lett.} \textbf{
  B733} (2014) 25--31,
  \href{http://dx.doi.org/10.1016/j.physletb.2014.04.007}{\doi{10.1016/j.physletb.2014.04.007}},
\href{http://www.arXiv.org/abs/1402.6965}{\texttt{ arXiv:1402.6965}}.

\bibitem{Finkbeiner:2014sja}
\hrefCMSnoop {} {D.~P. Finkbeiner and N.~Weiner, ``{An X-Ray Line from eXciting
  Dark Matter}'',}
\href{http://www.arXiv.org/abs/1402.6671}{\texttt{ arXiv:1402.6671}}.

\bibitem{Ishida:2014dlp}
\hrefCMSnoop {} {H.~Ishida, K.~S. Jeong, and F.~Takahashi, ``{7 keV sterile
  neutrino dark matter from split flavor mechanism}'',}
  \href{http://dx.doi.org/10.1016/j.physletb.2014.03.044}{\doi{10.1016/j.physletb.2014.03.044}},
\href{http://www.arXiv.org/abs/1402.5837}{\texttt{ arXiv:1402.5837}}.

\bibitem{Baek:2014poa}
\hrefCMSnoop {} {S.~Baek, P.~Ko, and W.-I. Park, ``{The 3.5 keV X-ray line
  signature from annihilating and decaying dark matter in Weinberg model}'',}
\href{http://www.arXiv.org/abs/1405.3730}{\texttt{ arXiv:1405.3730}}.

\bibitem{Nakayama:2014rra}
\hrefCMSnoop {} {K.~Nakayama, F.~Takahashi, and T.~T. Yanagida, ``{Extra light
  fermions in $E_6$-inspired models and the 3.5 keV X-ray line signal}'',}
\href{http://www.arXiv.org/abs/1405.4670}{\texttt{ arXiv:1405.4670}}.

\bibitem{Chakraborty:2014tma}
\hrefCMSnoop {} {S.~Chakraborty, D.~K. Ghosh, and S.~Roy, ``{7 keV Sterile
  neutrino dark matter in $U(1)_R-$ lepton number model}'',}
\href{http://www.arXiv.org/abs/1405.6967}{\texttt{ arXiv:1405.6967}}.

\bibitem{Ibrahim:2008gg}
\hrefCMSnoop {} {T.~Ibrahim and P.~Nath, ``{An MSSM Extension with a Mirror
  Fourth Generation, Neutrino Magnetic Moments and LHC Signatures}'',} \textit{
  Phys.Rev.} \textbf{ D78} (2008) 075013,
  \href{http://dx.doi.org/10.1103/PhysRevD.78.075013}{\doi{10.1103/PhysRevD.78.075013}},
\href{http://www.arXiv.org/abs/0806.3880}{\texttt{ arXiv:0806.3880}}.

\bibitem{Gunion:1984yn}
\hrefCMSnoop {} {J.~Gunion and H.~E. Haber, ``{Higgs Bosons in Supersymmetric
  Models. 1.}'',} \textit{ Nucl.Phys.} \textbf{ B272} (1986) 1,
\href{http://dx.doi.org/10.1016/0550-3213(86)90340-8}{\doi{10.1016/0550-3213(86)90340-8}}.

\end{thebibliography}
\providecommand{\href}[2]{#2}\begingroup\raggedright
\endgroup

\end{document}